\documentclass[11pt]{article}
\usepackage{epsfig,dsfont}

\begin{document}
\sffamily

\thispagestyle{empty}
\vspace*{15mm}

\begin{center}

{\LARGE Adjoint quarks and fermionic boundary conditions} 
\vskip20mm
Erek Bilgici$^1$, Christof Gattringer$^1$, Ernst-Michael Ilgenfritz$^{1,2,3}$, Axel Maas$^1$
\vskip10mm
$^1$Institut f\"ur Physik, Karl-Franzens-Unversit\"at Graz,  \\
Universit\"atsplatz 5, A-8010 Graz, Austria \\
\vskip1mm
$^2$Institut f\"ur Physik, Humboldt-Universit\"at zu Berlin,\\
Newtonstra{\ss}e 15, D-12489 Berlin, Germany \\
\vskip1mm
$^3$Institut f\"ur Theoretische Physik, Universit\"at Heidelberg,\\
Philosophenweg 19, D-69120 Heidelberg, Germany
\end{center}
\vskip27mm

\begin{abstract}
We study quenched SU(2) lattice gauge theory with adjoint fermions in a  wide
range of temperatures. We focus on spectral quantities of the Dirac operator and
use the temporal fermionic boundary conditions as a tool to probe the system. We
determine the deconfinement temperature through the Polyakov loop, and the
chiral symmetry restoration temperature for adjoint fermions through the gap in
the Dirac spectrum. This chiral transition temperature is about four times
larger than the deconfinement temperature. In between the two transitions we
find that the system is characterized by a non-vanishing chiral condensate which
differs for periodic and anti-periodic fermion boundary conditions. Only for the
latter (physical) boundary conditions, the condensate vanishes at the chiral
transition. The behavior between the two transitions suggests that deconfinement
manifests itself as the onset of a dependence of spectral quantities of the
Dirac operator on boundary conditions. This picture is supported further by our
results for the dual chiral condensate.\\

To appear in JHEP.
\end{abstract}

\setcounter{page}0
\newpage
\noindent
{\Large 1. Introductory remarks}
\vskip4mm
\noindent
Confinement and chiral symmetry breaking are two outstanding properties of
Quantum Chromodynamics (QCD), shaping all of nuclear physics. The emergence of
both these phenomena depends on  non-perturbative mechanisms, but it is an open
question if and how the respective mechanisms are related, or what they are in
detail.  Each phenomenon is connected with a particular symmetry becoming broken
or restored in certain limits of the theory.

From the moment on, at which the existence of phase transitions in QCD was
realized, the question whether deconfinement and chiral symmetry restoration are
related to different transition temperatures, $T_{dec}$ and $T_{ch}$, was posed.
Because of the dual role of quarks, with quarks being confined on one hand, and
their role in the (chiral) hadron dynamics on the other hand, this question was
asked in the first instance about fermions in the fundamental representation.
For this case a consensus~\cite{somelatticerapport} based on lattice gauge theory
calculations has formed that, at least as long as no  finite baryonic chemical
potential $\mu$ is involved, the temperature driven  phase transition happens at
roughly the same temperature,  $T_{dec} \simeq T^{(f)}_{ch}$, where we use the
superscript $(f)$ to indicate the fundamental representation. \footnote{For most
quark masses, including likely the physical ones, there is a crossover instead
of a genuine phase transition~\cite{somelatticerapport} in full QCD. Thus, there
is no qualitative distinction of the low- and  high-temperature phase, despite
their historic names. In quenched QCD  ("gluodynamics"), however, a second or
first order phase transition exists.} 

Fermions, and thus implicitly also chiral symmetry, play a role also outside
QCD, e.g., for model building beyond the standard model~\cite{somereview}. In
particular in many of those theories~\cite{beyond_SM}, like supersymmetry and
technicolor, fermions in other representations appear, in particular adjoint
ones.  There is no {\it a priori} reason to expect the same transition
temperature for such fermion representations. 

Such gauge theories with adjoint fermions have been investigated since the 
early days of lattice simulations
\cite{Kogut83,Kogut85,Kogut87,Luetgemeier98,Luetgemeier99,Engels:2005te}. In
this case, it is well established that, at least for the gauge groups
investigated so far, the deconfinement temperature $T_{dec}$ and the chiral
restoration temperature $T^{(a)}_{ch}$ of adjoint quarks  do not coincide, the
latter being generally significantly larger than the former. 

Aside from the practical considerations of theories beyond the standard model,
this requires that any mechanism proposed as an explanation for the equality
$T_{dec} \simeq T^{(f)}_{ch}$ must at the same time provide an explanation for
the inequality $T_{dec} \neq T^{(a)}_{ch}$. Such attempts have been made
already. E.\ g., to capture the characteristic temperatures $T^{(r)}_{ch}$ of
chiral symmetry  restoration for some representation $r$,  a hypothetical
Casimir scaling law has been proposed \cite{Marciano,Raby}, $C_2^{(r)}
g^2(T^{(r)}_{ch}) = {\rm const} \approx 4$, and discussed
in the light of early lattice results in Refs.~\cite{Kogut83,Kogut85}. Here $C_2^{(r)}$ is the  eigenvalue
of the quadratic Casimir operator that characterizes the fermion  representation
$r$, and $g^2$ is the running coupling.

Out of the necessity to explain the difference it is also possible to construct
a virtue: The fact that for adjoint fermions the temperatures for deconfinement
and chiral symmetry restoration are different, makes such theories an important
testbed to study confinement and chiral symmetry breaking individually. In this
way one may hope to understand mechanisms responsible for the two phenomena and
to identify possible aspects shared by both.

This point of view is the motivation for the present work: In a series of recent
papers~\cite{grazregensburg,dualcondensate2008,soeldner,wipf,Fischer} the
question of a possible connection  between confinement and chiral symmetry
breaking has been attacked by constructing new combined observables which are
sensitive to both, confinement and chiral symmetry breaking. One example is the
"dual chiral condensate" \cite{dualcondensate2008}, which is obtained as the
first Fourier component of the chiral quark condensate with respect to a
generalized temporal boundary condition for the fermions. It may be shown
\cite{dualcondensate2008} that the dual chiral condensate for fundamental
fermions is a sum of generalized (i.e., non-straight) Polyakov loops and thus is
an order parameter for center symmetry and therefore for confinement (at least
in the quenched case). On the other hand, since it is built from the usual
chiral condensate, it is also sensitive to chiral symmetry breaking.

This observable can be used to characterize the deconfinement temperature
$T_{dec}$ as the one above which spectral quantities of the Dirac operator
(e.g., the chiral condensate) become sensitive to a change of the temporal
fermion boundary conditions \cite{dualcondensate2008}. The origin of the tie
to chiral symmetry is the density of Dirac eigenvalues at the origin, due to
the Banks-Casher relation \cite{BanksCasher}. This characterization is further
underlined by the fact that both the chiral condensate and the gap in the
Dirac spectrum above $T_{dec}$ are quantities that depend on the fermionic
boundary conditions (relative to the phase of the Polyakov loop)
\cite{Z3_sectors,gapboundary,Luschevskaya}.  For the case of SU(2) Yang-Mills
theory the detailed circumstances suggest the following microscopic
explanation~\cite{Martemyanov} for that dependence: For periodic boundary
conditions low-lying  modes exist that are localized on "light dyons",  
whereas "heavy dyons" are  suppressed. The latter would otherwise be carriers
of low-lying modes under anti-periodic boundary conditions. In this picture,
the different abundance of light and heavy dyons in turn results from the
non-vanishing fundamental Polyakov loop.

The central motivation for the present investigation is the question to what
extent the interrelations between deconfinement, chiral symmetry restoration and
the fermionic boundary conditions carry over to the case of fermions in the
adjoint representation of the gauge group. Of particular interest is the
behavior in the intermediate phase, i.e., at temperatures  $T_{dec} (\simeq
T_{ch}^{(f)}) \leq T \leq T_{ch}^{(a)}$. For this range we will show that the
condensate is still finite and no spectral gap has opened, but the fermionic
quantities do already feel the boundary conditions. Concerning the dual chiral
condensate we will establish that it is sensitive to both the deconfinement and
chiral restoration transitions.

\vskip7mm
\noindent
{\Large 2. Setup of the calculation}
\vskip4mm
\noindent
In our analysis we study quenched SU(2) configurations generated with the
Symanzik improved gauge action \cite{gaugeact} using the fundamental
representation. We explore a wide range of inverse couplings $\beta$, between
$\beta = 2.5$ and $\beta = 4.6$, increasing $\beta$ in steps of $\Delta \beta =
0.1$. Using Metropolis updates, for each value
of $\beta$ we generate 100 configurations in each of our ensembles on two
volumes, $N^3 \times N_T = 10^3 \times 4$ and $12^3 \times 4$. 

The fundamental gauge links $U_\mu(x)$ are converted to the adjoint
representation
\begin{equation}
U_\mu^{adj}(x)_{ab} \; \equiv \; 
\frac{1}{2} \, 
\mbox{Tr} \, [ \, \sigma^a \, U_\mu(x)^\dagger \, \sigma^b \, U_\mu(x) \, ] 
\; ,
\end{equation}
where $\sigma^a,\, a = 1,2,3$ are the Pauli matrices. The adjoint links are used
in the massless staggered lattice Dirac operator (we set the lattice spacing to
$a = 1$)
\begin{equation}
D(x,y) \; = \; \sum_\mu \, \eta_\mu(x) \, [ \; U_\mu^{adj}(x) \, \delta_{x+\hat{\mu},y} 
\, - \, U_\mu^{adj}(x-\hat{\mu})^\dagger \, \delta_{x-\hat{\mu},y} \; ] \; ,
\end{equation}
where $\eta_\mu(x)$ is the staggered sign function $\eta_\mu(x) = 
\prod_{\nu=1}^{\mu-1}(-1)^{x_\nu}$. 

For the staggered lattice Dirac operator we evaluate  complete eigenvalue
spectra using a parallel implementation of standard linear algebra routines. The
staggered Dirac operator is anti-hermitean and consequently the eigenvalues
$\lambda_j$ are purely imaginary. The eigenvalues for the Dirac operator with
mass $m$ are then given by $\lambda_j + m$.

In our analysis we systematically explore the role of the temporal fermionic
boundary conditions, which may be written as
\begin{equation}
\psi(\vec{x},N_T) \; = \; e^{i\varphi} \, \psi(\vec{x},0) \; ,
\end{equation}
where the "boundary angle" $\varphi$ parameterizes the boundary condition. A
value of $\varphi = \pi$ corresponds to the usual anti-periodic boundary
conditions. However, here in addition we explore also periodic and more general
boundary conditions, and the boundary angle $\varphi$ is considered as an
additional parameter to probe the system. Furthermore, for the construction of
the aforementioned dual chiral condensate we need a Fourier integral over
$\varphi$ which is approximated by using altogether 8 values of $\varphi$ in the
interval $[0,2\pi)$.  To be specific, we compute complete Dirac spectra for the
two boundary conditions $\varphi = 0$ and $\varphi = \pi$ for all 100
configurations in our ensembles, while spectra for the additional values 
$\varphi = \pi/4, \, \pi/2, \, 3\pi/4,  \, ... \, $ needed for the dual chiral
condensate were evaluated for subensembles consisting of only 20 configurations
for each volume and $\beta$. For completeness we remark, that all other boundary
conditions, i.e., the spatial fermionic boundary conditions and the boundary
conditions for the gauge fields, were kept periodic. 

\begin{figure}[p!]
\begin{center}
\includegraphics[width=125mm,clip]{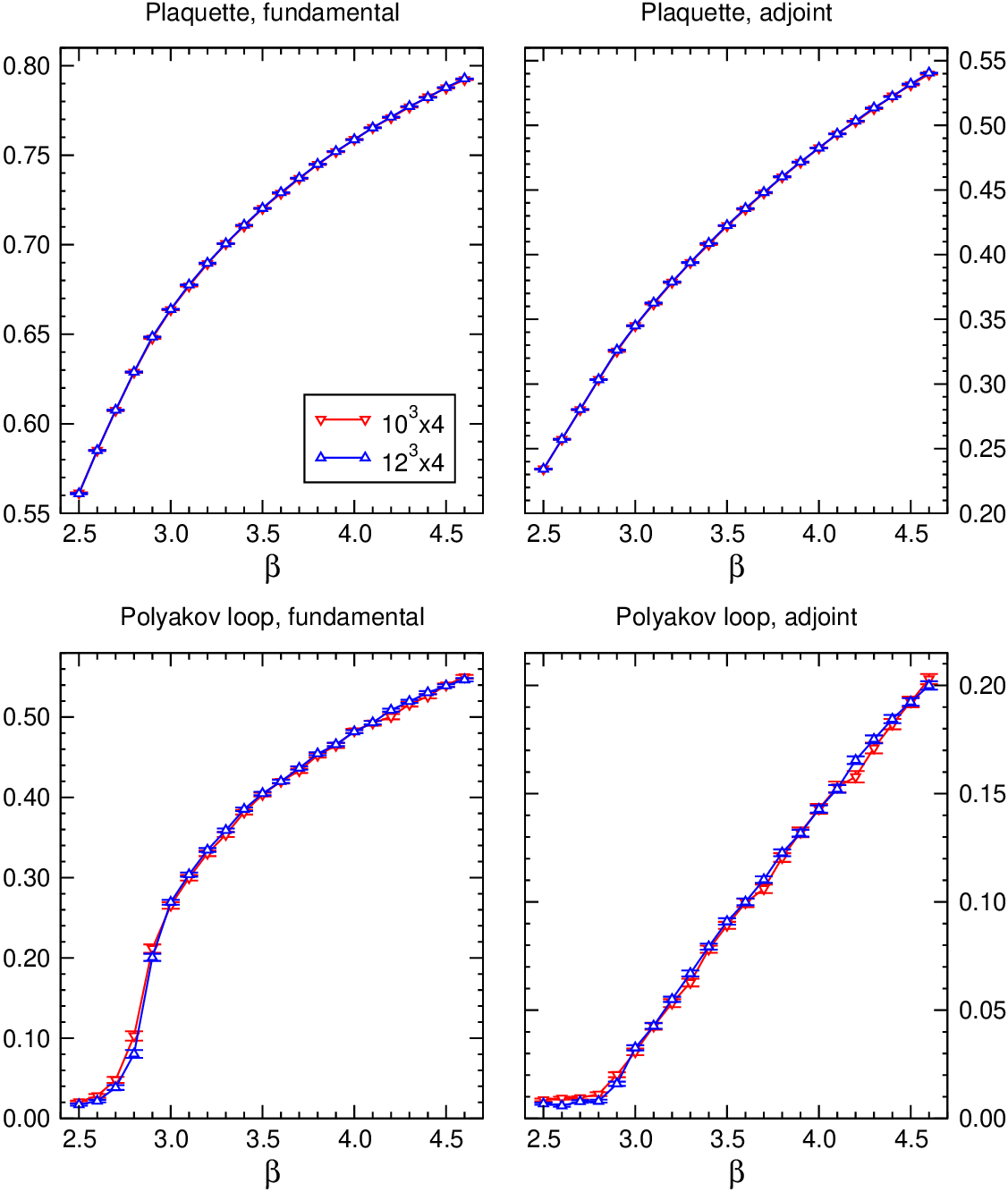} 
\end{center}
\caption{Gluonic observables for our quenched gauge ensembles. We show the
fundamental and adjoint plaquette expectation values as a function of the
inverse coupling $\beta$ (top row of plots), and the fundamental and adjoint
Polyakov loops (bottom row). Results for both volumes, $10^3 \times 4$ and $12^3
\times 4$ are displayed.} 
\label{gluonicobservables} 
\end{figure}

\vskip7mm
\noindent
{\Large 3. Plaquette and Polyakov loops}
\vskip4mm
\noindent
We begin our discussion of the numerical results with purely gluonic quantities,
the plaquette expectation values and the (spatially averaged) Polyakov loops in
both the fundamental and the adjoint representations.

In Fig.~\ref{gluonicobservables} we compare the fundamental and adjoint
plaquette expectation values (top row) and the fundamental and adjoint  Polyakov
loops (bottom row) plotted as a function of the inverse gauge coupling $\beta$.
The fundamental Polyakov loop can be used to determine the critical inverse
gauge coupling where we observe the deconfinement transition  on our lattice
with $N_T = 4$, at a value of $\beta_ {dec} = 2.8$. This
corresponds\footnote{For a string tension of $\sqrt\sigma=440$ MeV, using
$a^2\sigma$ results from \cite{Bornyakov:2005iy}.} to about $T=215$ MeV, and
thus is very far from the continuum and infinite-volume limit of about 300 MeV
\cite{Fingberg:1992ju}. Our plots clearly indicate that a comparison of
the data for the two different volumes, $10^3 \times 4$ and $12^3 \times 4$,
reveals only very small finite volume effects.

It is a remarkable fact that also the adjoint Polyakov loop shows a changing
behavior at the onset of the deconfinement transition.  Since it is invariant
under center transformations, there is no a-priori reason for this. This behavior
will be one contribution to the sensitivity of the adjoint dual chiral
condensate discussed below. However, the impact on the adjoint Polyakov loop by
the breaking of center symmetry could be spurious: Adjoint fermions can be
screened by a single gluon. Thus, even for static adjoint quarks string breaking
occurs for all temperatures \cite{Bali:2000gf}, and there is no deconfinement in
the same sense as there is none for full QCD. Hence, in the infinite-volume and
continuum limits, the adjoint Polyakov loop is non-zero in all phases. It is not
an order parameter for center symmetry.

The fact that on a finite lattice an imprint of the deconfinement transition
still exists has also been observed in G$_2$ Yang-Mills theory
\cite{g2paper,Pepe:2006er}, and is thus not surprising. Still, this demands
caution in the interpretation of adjoint quantities, and the value of the
adjoint Polyakov loop has to be interpreted rather as a lattice artifact than as
a signal, as long as it cannot be unambiguously established that other effects
drive its modification.

\vskip7mm
\noindent
{\Large 4. Spectral gap and chiral condensate}
\vskip4mm
\noindent
Let us now come to fermionic observables related to chiral symmetry breaking and
its restoration. In this respect an important result is the Banks-Casher formula
\cite{BanksCasher} which relates the chiral condensate to the density $\rho$  of
Dirac eigenvalues at the origin, 
\begin{equation}
\langle \, \overline{\psi} \, \psi \, \rangle \; = \; 
- \, \pi \, \rho(0) \; .
\label{bacaformula}
\end{equation}
This result is independent of the gauge group and its representation. As long as
chiral symmetry is broken we thus expect that the eigenvalues of the Dirac
operator extend all the way to the origin and build up a non-vanishing density 
$\rho(0)$ there. As one crosses the critical temperature, the chiral condensate
vanishes and so must $\rho(0)$. At least on finite spatial volumes one observes
the opening of a gap in the spectrum at a corresponding critical coupling
$\beta_{ch}^{(a)}$.\footnote{With the currently available results it cannot be
excluded that in the thermodynamic limit the gap closes such that the
eigenvalues extend all the way to the origin, but still have a vanishing density
$\rho(0)$ \cite{tamas}.} 

\begin{figure}[t!]
\begin{center}
\hspace*{-3mm}
\includegraphics[width=125mm,clip]{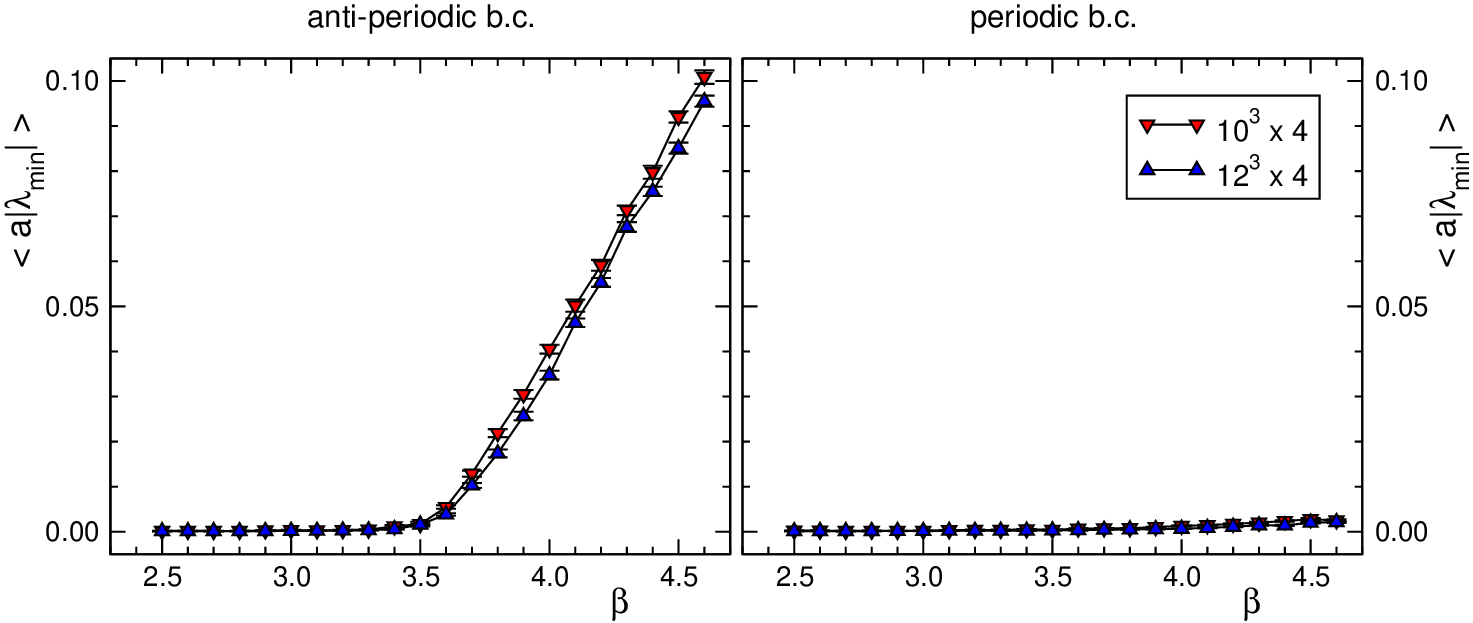}
\end{center}
\caption{Spectral gap in lattice units of the Dirac operator for adjoint fermions 
as a function of the inverse coupling $\beta$. We 
compare anti-periodic temporal boundary conditions for the fermions (lhs.~plot)
to periodic boundary conditions (rhs.). Results for both volumes, 
$10^3 \times 4$ and $12^3 \times 4$, are displayed.} 
\label{spectralgap} 
\end{figure}

For the case of adjoint SU(2) we expect that chiral symmetry is restored at a
higher temperature than the one where we observe deconfinement and thus expect
that the spectral gap remains closed beyond $\beta_{dec} = 2.8$. This is exactly
what we observe in the lhs.\ plot of Fig.~\ref{spectralgap}. Using lattice units
we show the spectral gap defined as the expectation value $\langle a
|\lambda_{min}| \rangle$ of the smallest eigenvalue as a function of $\beta$.
The plot clearly shows that the gap remains closed above $\beta_{dec} = 2.8$ all
the way up to $\beta_{ch}^{(a)} = 3.6$, corresponding to about 870 MeV, where it
starts to open (again we use the superscript $(a)$ to denote the critical
$\beta$ for the adjoint representation). We observe that the discrepancy between
the two volumes which are accessible to us, $10^3 \times 4$ and $12^3 \times 4$,
is small with a light trend towards a smaller gap for the larger spatial
volume. 

One may compare the two critical inverse couplings $\beta_{dec} = 2.8$ and 
$\beta_{ch}^{(a)} = 3.6$ also in terms of temperatures. One finds that the
deconfinement temperature and the temperature for chiral symmetry restoration
behave as
\begin{equation}
T_{ch}^{(a)} \; \simeq \; 4(1) \, T_{dec} \; . \; \; 
\end{equation}
\noindent 
The error is a rough estimate based on the discrepancy of the determined
deconfinement temperature and the known infinite-volume continuum value. Still,
both transitions are different. However, they are much closer than in the case
of (dynamical) SU(3) QCD, where they differ by a factor of 7.8(2)
\cite{Luetgemeier99,Engels:2005te}.

Let us now discuss the rhs.~plot of Fig.~\ref{spectralgap} which differs from
the lhs.~by the use of periodic temporal boundary conditions for the fermions
instead of the canonical  anti-periodic choice. Obviously the spectral gap
remains closed when the periodic temporal boundary conditions are used for the
fermions. This behavior is in agreement with what was found also for the gauge
groups SU(3), SU(2) and G$_2$ in the fundamental representation
\cite{gapboundary,Luschevskaya,g2paper}. 

\begin{figure}[t]
\begin{center}
\hspace*{-8mm}
\includegraphics[width=120mm,clip]{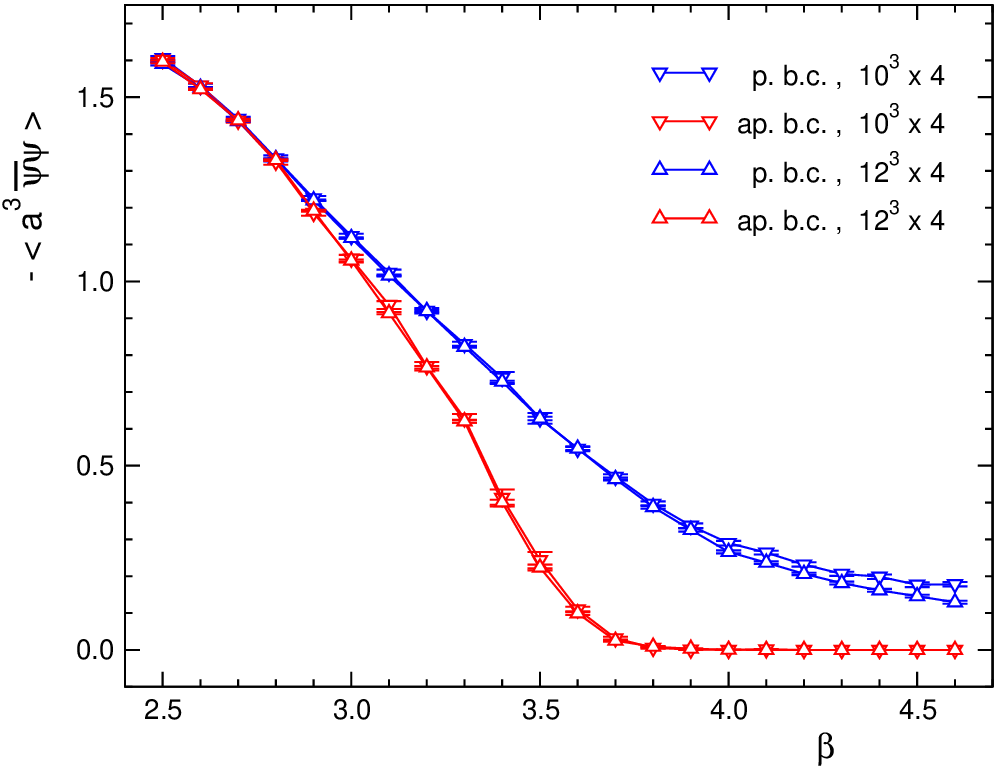}
\end{center}
\caption{The chiral condensate in lattice units as a function of the inverse
gauge coupling $\beta$. We show the results for both volumes,  $10^3 \times 4$
and $12^3 \times 4$, and compare periodic and anti-periodic temporal boundary
conditions for the fermions. Note that this is the unrenormalized chiral condensate.} 
\label{condensate} 
\end{figure}

The next step is to explore the dependence of the chiral condensate on the
temperature and the boundary conditions directly. For this purpose, the
condensate is determined in the same way as in  \cite{g2paper}, using both methods. In
Fig.~\ref{condensate} we plot the chiral condensate in lattice units as a
function of the inverse gauge coupling $\beta$ and compare periodic (upper two
curves) and anti-periodic (lower two curves) temporal boundary conditions   for
the fermions. Triangles are used for the larger $12^3 \times 4$  lattice,
while the smaller $10^3 \times 4$ lattice is represented by upside-down triangles. For
the anti-periodic boundary conditions we observe that the condensate remains
finite up to about $\beta_{ch}^{(a)} = 3.6$, the critical value where we
observed the opening of the spectral gap, and vanishes for larger $\beta$. The
situation is different for the condensate with periodic boundary conditions
where we find that the condensate remains finite above $\beta_{ch}^{(a)} = 3.6$,
as could be already expected from the fact that no spectral gap appears 
(compare Fig.~\ref{spectralgap}). Again we find that the results for the two
volumes essentially fall on top of each other -- only for the periodic case at
the largest values of $\beta$ we observe sizable finite volume effects. However, at such large values of $\beta$ the spatial volume becomes so small that the results can be taken only as indicative. Still, since the major effects investigated, i.\ e., the chiral and deconfinement phase transitions, occur at $\beta<4$, where no such effects are visible, the conclusions are likely not affected qualitatively by this limitation. It would be necessary to use significantly larger volumes to obtain a better systematic accuracy.

Let us finally stress an important aspect of our results for the chiral
condensate: By comparing the data for periodic and anti-periodic boundary 
conditions at the smallest values of $\beta$ we find that the results for the
condensate fall on top of each other. This is true up to the value of $\beta =
2.8 = \beta_{dec}$. Beyond the value of the deconfinement transition we observe
that the results for the condensate at periodic and anti-periodic boundary
conditions start to differ. The condensate for the anti-periodic case begins to
drop relative to the periodic data until it reaches zero near $\beta_{ch}^{(a)}
= 3.6$. In between the two transitions we observe a finite chiral condensate for
both boundary conditions but the values differ. Note that the effect is
significantly stronger than the systematic finite-volume errors.

This finding  underlines  the characterization of deconfinement and chiral
symmetry restoration given in  \cite{dualcondensate2008}: The deconfinement
transition is characterized by the onset of a dependence of the Dirac spectrum
on the fermionic boundary conditions. Chiral symmetry restoration is seen only
for the physical anti-periodic boundary conditions and, according to the
Banks-Casher formula, is manifest through a vanishing spectral density at the
origin.

\vskip7mm
\noindent
{\Large 5. The dual chiral condensate}
\vskip4mm
\noindent     
In a series of papers \cite{grazregensburg,dualcondensate2008,wipf} observables
were developed that are sensitive to both, chiral symmetry and confinement. One
such observable is the dual chiral condensate $\Sigma_1$ which is defined as the
first Fourier component of the chiral condensate with respect to the fermionic
temporal boundary condition \cite{dualcondensate2008},
\begin{equation}
\Sigma_1 \; = \; - \, \frac{1}{2\pi} \int_0^{2\pi} \!\!\!
d \varphi \; e^{- i \varphi} \,
\big\langle \,\overline{\psi} \psi \, \big\rangle_m^{(\varphi)} \; = \;
\frac{1}{2\pi V} \int_0^{2\pi} \!\!\!
d \varphi \,\sum_j \frac{e^{- i \varphi}}{\lambda_j^{(\varphi)} + m} \; .
\label{dualconddef}
\end{equation}
In our notation the superscript $(\varphi)$ indicates which fermionic boundary
condition is used. In the second step of (\ref{dualconddef}) we have inserted
the spectral sum for the Dirac operator. The integral over the boundary angle is
approximated with 8 values of $\varphi$ in the interval $[0, 2\pi)$ using the
Simpson rule. This procedure was shown to give rise to uncertainties for the
numerical integral in the one percent range \cite{dualcondensate2008}, in case
of the smooth dependence on the boundary angle observed here.

The dual chiral condensate $\Sigma_1$ may be viewed as a collection of
generalized Polyakov loops: Like any other gauge invariant quantity on the
lattice the scalar expectation value $\langle \overline{\psi} \psi \rangle_m$
can be expressed as a collection of closed loops on the lattice which are
dressed with link variables. These loops may be distinguished by their winding
number around the compactified time direction and the Fourier transformation
with respect to the boundary angle $\varphi$ in (\ref{dualconddef}) projects to
the equivalence class of loops that wind once. Consequently these loops that
build up $\Sigma_1$ transform under center transformations like the conventional
straight Polyakov loop. Thus in the quenched theory with (current) fermions in
the fundamental representation $\Sigma_1$ serves as order parameter for center
symmetry and thus for confinement. In addition, for small enough quark mass $m$
the observable becomes sensitive to chiral symmetry breaking since it is derived
from the conventional chiral condensate. In the limit of large quark mass longer
loops are suppressed and $\Sigma_1$ approaches the conventional straight
Polyakov loop (with a different overall normalization). 

\begin{figure}[t]
\begin{center}
\hspace*{-2mm}
\includegraphics[width=125mm,clip]{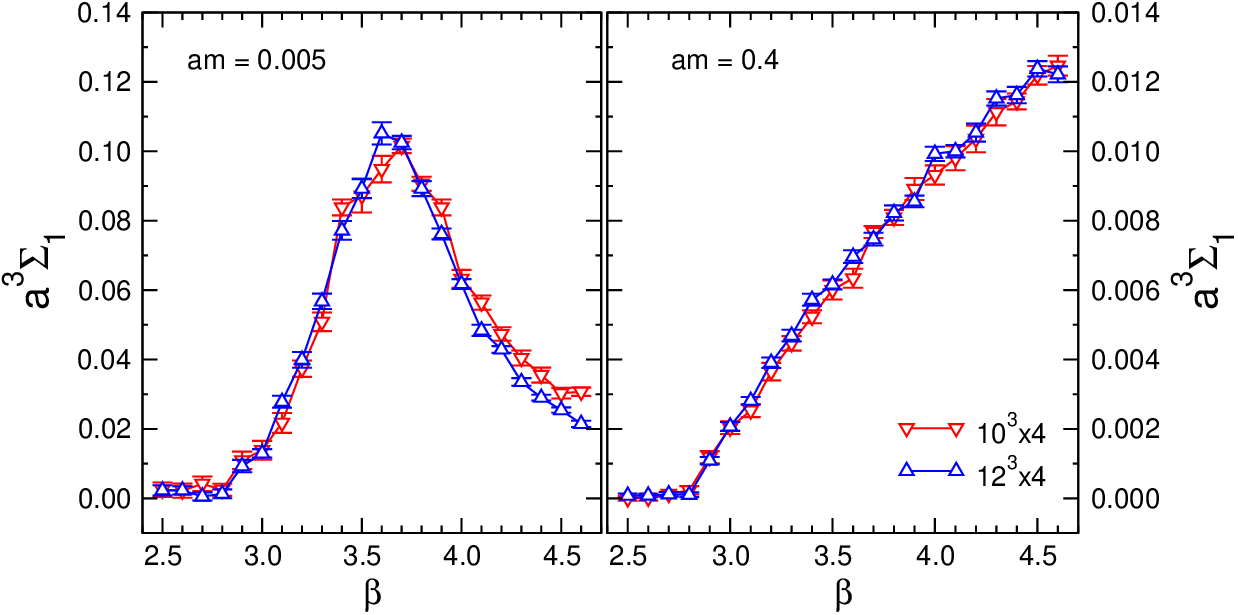}
\end{center}
\caption{The dual chiral condensate $\Sigma_1$ in lattice units as a function of
the inverse gauge coupling $\beta$. We show the results for both volumes, $10^3
\times 4$ and $12^3 \times 4$, and compare two different quark masses.} 
\label{dualcondensate} 
\end{figure}

Since for adjoint fermions deconfinement and chiral symmetry restoration appear
at different temperatures, it is an interesting question how the adjoint dual
chiral condensate behaves in this situation. Furthermore, as discussed above,
the adjoint Polyakov loops from which the adjoint dual chiral condensate is
constructed, are invariant under center transformations. The behavior of the
adjoint dual chiral condensate is therefore much harder to predict on general
grounds than in the fundamental case.

In Fig.~\ref{dualcondensate} we show our results for the dual chiral condensate as a
function of the inverse gauge coupling $\beta$. In the lhs.\ plot we consider a
situation which corresponds essentially to the chiral limit ($am = 0.005$),
while on the rhs.\ we use a  rather large quark mass ($am = 0.4$) where
$\Sigma_1$  is expected to behave similar to the conventional straight Polyakov
loop. The plots show clearly that $\Sigma_1$ starts to rise at the deconfinement
transition at $\beta_{dec} = 2.8$. Since the dual chiral condensate is the first
Fourier component of the condensate with respect to the fermionic boundary
conditions, its behavior supports the characterization of the deconfinement
transition as the onset of dependence of fermionic quantities on the fermionic
boundary conditions.

Comparison of the two plots in Fig.~\ref{dualcondensate} shows that only for the
small quark mass $\Sigma_1$ also the chiral symmetry restoration at
$\beta_{ch}^{(a)} = 3.6$ is resolvable. In the lhs.\ plot we observe a maximum
of $\Sigma_1$ at this coupling. Beyond this value we find a decreasing behavior.
In the rhs.\ plot, where the rather large mass $am = 0.4$ was used, we find no
signal at $\beta_{ch}^{(a)} = 3.6$ and, as expected, $\Sigma_1$ behaves similar
to the conventional straight Polyakov loop in the adjoint representation
(compare Fig.\ \ref{gluonicobservables}), i.e., displays a monotonically rising
behavior beyond $\beta_{dec} = 2.8$.

\vskip7mm
\noindent
{\Large 6. Summary and discussion}
\vskip5mm
\noindent    
In this paper we have revisited the phenomenon of different deconfinement and
chiral symmetry restoration temperatures of gauge theories coupled to adjoint
fermions. In our study of quenched SU(2) gauge configurations we have focused on
analyzing a set of observables related to the spectrum of the lattice Dirac
operator. An important tool in this analysis was the use of generalized temporal
fermionic boundary conditions. The corresponding boundary angle serves as an
additional parameter to probe the system.

We confirm that for adjoint fermions the deconfinement transition (determined by
the Polyakov loop expectation value) and the chiral symmetry restoration
(identified by the opening of a gap in the Dirac spectrum), are different with
$T_{ch}^{(a)} = 4(1) \, T_{dec}$. We find that also fermionic quantities are
affected by the deconfinement transition, in particular at the deconfinement
temperature a dependence of the chiral condensate on the fermionic boundary
condition becomes manifest. Between the two transitions the system is
characterized by a chiral condensate which differs for different boundary
conditions, but for the physical anti-periodic boundary conditions still has not
reached zero. This happens at the second transition $T_{ch}^{(a)}$ where chiral
symmetry is restored, i.e., the chiral condensate finally vanishes. However, 
we do not find any indications of this being related to a thermodynamic
phase transition of the pure gauge system, at least in any gluonic
observable we have investigated. This would be in line with observations for the
dynamical case \cite{Luetgemeier99,Engels:2005te}. The only affected quantities 
are fermionic ones: For periodic temporal fermion boundary conditions the 
condensate remains finite for all temperatures (i.e., all gauge couplings) 
we considered, though the systematic uncertainty increases quickly for $\beta>4$. Finally we find that the
dual chiral condensate indeed sees both transitions, thus providing support that
this observable is sensitive to both confinement and chiral symmetry breaking.
Still, a careful study of the thermodynamic limit is mandatory for a firm
conclusion, in particular at temperatures beyond the chiral phase transition.

An interesting question is how the picture changes when dynamical adjoint quarks
would be used. From dynamical SU(3) QCD
studies~\cite{Luetgemeier99,Engels:2005te} it is known that the back-reaction of
the fermions on the gluon field is minor for large masses in the sense that the
chiral phase transition at $T^{(a)}_{ch} \approx 7.8(2) \, T_{dec}$ has no
strong effect on the remaining observables. On the other side it is known that
the deconfining  phase transition is mainly unaffected by the presence of
dynamical adjoint fermions, the chiral symmetry of which is broken for all $T <
T^{(a)}_{ch}$. However, in contrast to the pure SU(3) Yang-Mills theory, the
deconfining phase transition is of strong first order. Nonetheless, we
expect that the picture developed here, i.e., an intermediate phase where 
fermionic quantities do already depend on the temporal fermionic boundary 
conditions but the chiral condensate is still
non-vanishing, carries over to the full dynamical theory, since this
effect should not be affected by the order of the phase transition. 

Our analysis has increased the amount of known phenomenological facts about
systems with deconfinement and chiral symmetry restoration transitions. In
particular the role of the fermionic boundary conditions, which have become an
important issue in recent years, was clarified for a system with adjoint
fermions. It is obvious that any future microscopic explanation of the
deconfinement and chiral symmetry restoration transitions will have to describe
the dependence on boundary conditions correctly.

A particular highlight of the results is that, since the calculations have been
quenched, all the mechanisms usually associated with chiral restoration, like
modification of topological properties, cannot be responsible for the
restoration of the adjoint chiral symmetry: All of these effects occur at
$T_{dec}$, unmodified in the quenched calculation. The dynamical origin of
adjoint dynamics is therefore fundamentally and qualitatively different from the
one for fundamental dynamics. Especially, as all dynamics driving fundamental
chiral symmetry breaking cease at a quarter of the temperature $T^{(a)}_{ch}$,
they alone cannot provide adjoint chiral symmetry breaking. In particular, this
implies a very strong adjoint-quark-gluon dynamics in the high-temperature phase
to keep chiral symmetry broken, but since everything is quenched, this implies
strong gluon dynamics even at $4\, T_{dec}$ (or nearly $8\, T_{dec}$ for SU(3)).

\vskip7mm
\noindent
{\Large Acknowledgments}
\vskip4mm
\noindent
We thank Falk Bruckmann, Tamas Kovacs, Christian Lang and Andreas Wipf  for
interesting discussions during the course of this work. Erek Bilgici was
supported by the FWF DK W 1203, Christof Gattringer partly by FWF grant number 
P20330-N16, and Axel Maas by FWF grant
number M1099-16. The numerical analysis was done on the clusters at ZID,
University of Graz. E.-M.I.\ gratefully acknowledges the guest position at  the
University of Graz, where this study was begun, and the interims position at
University of Heidelberg that he presently holds.

\end{document}